\documentstyle[11pt]{article}
\begin{document}
\hyphenation{gra-vi-ta-tio-nal pa-ral-lel en-coun-ter en-coun-ters}
\title{Gravitational Shapiro phase shift 
on pulsars' period to detect Dark Matter}
\author{D.~Fargion$^1$ and
R.~Conversano\\
$^1$Physics Department, University of Rome, P.le Aldo Moro 2, \\
00185 Rome, Italy ; INFN - ROMA1}
\maketitle
\begin{abstract}
The Shapiro Phase Shift on Pulsars ({\bf SPSP}) influence at low , but 
detectable level the pulsars' ({\bf PSR}) periods and their derivatives. This 
effect may 
be already written in PSRs' time derivative deviation and in particular it 
may explain the presence of a few negative pulse derivatives among 558 known 
PSRs. We describe in detail the phenomenon and predict that the {\bf SPSP} 
may play a role in $\dot{P}$ deviations for just a few PSRs a year ($\, 
\sim 0.5 \div 5 \, ev\,yr^{-1}\,$) 
at a$\,\dot{P} \sim \left( 10^{-14} \div 10^{-
15} \right) s \, s^{-1}\,$ level , leading to a plausible explanation 
to the puzzling negative pulse derivative of few PSRs' periods. 
\end{abstract}
 
\section{introduction}
The search for gravitational microlensing events of stellar sources produced 
by MAssive Compact Halo Objects ({\bf MACHOs}) provides an interesting but 
approximated tool for the study of dark matter in our galaxy. Although several 
events of achromatic magnification have been already observed , they have not 
let the MACHOs' parameters \mbox{(mass ,} transversal velocity ...) be known 
independently. Comparisons of the observed optical depths with those provided 
by theoretical galactic models have just suggested that there must be more 
matter toward the galactic centre than in a galactic halo , making more 
probable that our galaxy is a barred one. Therefore the real nature of the 
deflector that produces an observed event remains still unclear.\\
Unfortunately the possibility of studying interference between the two beams 
produced by a deflector fails , due to the inchoerence of the stellar source. 
Nevertheless the temporal phase shift produced by the deflector on `` 
monochromatic and coherent '' PSRs' signals , the Shapiro \mbox{effect ,} 
might 
provide another kind of investigation on the MACHO's parameters. The extremely 
high precision the PSRs' period can be known , and the explicit dependence 
of the time delay on the deflector's mass , may offer a deeper tool for the 
measurement of both the deflector's mass and impact parameter. An event of 
temporal phase gravitational shift , due to the massive object crossing the 
ray's path between the PSR and the observer , produces the same 
phenomenon of displacing away and back the PSR on distances of the order of 
the Schwarzschild radius of the deflector. No Doppler shift (due to the 
gravitational pulling of a nearby crossing star) may lead to a similar 
forth-and-back phase shift , but it may only induce a one-time phase 
stretching. Doppler effects due to binary systems introduce a sinousoidal
perturbation on residual PSRs' signal of the same order of magnitude but with 
a different time signature and a lower probability.\\
A program of detection would require a continuous monitoring of the PSRs' 
period because every event is a unique one and leaves no trace ; however it 
may hold for a period long enough to be observed.\\
We analyze the properties of this phenomenon by using the simple laws of 
gravitational lensing as found in literature. While the detectable 
gravitational magnification lensing needs millions of star sources (in order 
to fill up optical depths of $\sim 10^{-7}$) , in our framework the Shapiro 
phase sensitivity (due to the logarithmic behaviour on the impact parameter) 
is much larger. This 
leads to larger detectable impact parameters , and so , larger optical depths 
(at least up to the needed $\tau \sim 10^{-2}$ for the phenomenon to be 
observable with hundreds of PSR sources).\\
A similar effect between the two images produced by a microlensing event was 
already introduced \cite{krauss} and it seems to be already observed on the 
PSR B0525+21 \cite{larch} : it is our belief that the associated optical 
depth be too low (see note at proof).

\section{TIME DELAY VARIABILITY}
Consider the gravitational lens \cite{Einstein} due to a MACHO 
passing by the line-of-sight of a PSR with transversal 
velocity $v_{\perp}$ at a distance $D_{d}$ from the observer. Although two 
images are produced by the lensing of the PSR , the secondary image is 
observable only when the MACHO enters within the microlensing tube in general 
defined by the threshold parameter $\;u = 1\;$ \cite{pacz,pacz2,griest} ; 
this means that the 
probability the secondary image be luminous enough is too low to take care of 
it , even if we assume at least $\sim 500$ PSRs. Therefore 
from now on we shall consider only the primary image being observable.\\
The formula approximated (see the exact one in appendix A) that gives the time 
delay of the primary image's 
lensed ray relative to the undeflected ray , very similar to the 
Shapiro one \cite{sha} , depends on both a geometrical and a gravitational 
term :
\[
\Delta t \,= \, \Delta t_{geo} + \left[ \Delta t_{grav} \right] \; = \; 
\frac{r_{S}}{4\,c} \; \left( \sqrt{u^2\,+\,4\,} \, - \, u \right)^2 \; + 
\]
\begin{equation}         
\; + \left[ \frac{r_{S}}{c}\,\log\left( \frac{8\,D_{s}}{r_{S}}  
\right) \, - \, \frac{2\,r_{S}}{c}\,\log\left( \sqrt{u^2 + 4\, } +
u  \right) \right]
\end{equation}
where $u$ is the adimensional impact parameter of the MACHO from the 
line-of-sight of the PSR in units of the Einstein radius : 
$\;u = b \,/\, R_{E}\;$ (see figure 1) , where
{\small
\begin{equation}
R_{E} \, = \, \sqrt{2 \, r_{S} \, \frac{D_{d} \, D_{ds}}{D_{s}} \,} = 
4.84 \cdot 10^{13} \; \sqrt{\left[ \frac{M}{M_{\odot}} \right] 
\left[\frac{D_{s}}{4 Kpc}\right] \left[\frac{x_{d}}{0.5}\right] 
\left[ 1 - \left( \frac{x_{d}}{0.5} \right) \right]\,} \,\;  cm.
\end{equation}
}
$r_{S}$ is the Schwarzschild radius of the MACHO , $D_{d}$ , $D_{s}$ are 
respectively the observer-deflector and observer-source distances and 
$D_{ds}$ 
the deflector-source distance. In the Einstein radius formula the 
substitutions $\,x_{d} = D_{d} / D_{s}\,$ and $\,1 - x_{d} = D_{ds} / 
D_{s}\,$
have been made. $D_{s}$ is expressed in characteristic PSRs' distance 
units.\\
The geometrical term decreases very 
rapidly with the increase of $u$ and it does not play any role as long as 
$\,u \gg 1\,$. The gravitational terms (a first one constant and a second 
one variable) account for 
most of the gravitational effect along the ray's path. The constant in the 
gravitational term underlines the positive behaviour of the time delay. 
The logarithmic 
divergence for $\;u R_{E} \gg D_{d} \, , \, D_{ds}\;$ , i. e. for \mbox{$\;u 
\rightarrow \infty$ ,} is fictious ; the time delay vanishes as 
{\small $\frac{r_{S}}{c}\,\frac{D_{s}}{b}$} , as shown in appendix A.
As the formula shows , 
the time delay depends explicitly on the mass of the MACHO because of the 
presence of $r_{S}$ ; so it gives more transparent and accurate informations 
on the lensing dark object than the measurement of any gravitational lensing 
magnification does ($\propto 1 / u$) at small $u$. Equation (1) is 
valid nearly always (as long as \mbox{$u \ll \sqrt{D_{s}/(2\,r_{S})} \leq 
10^8$ ,} for 
a characteristic solar mass) , namely for any realistic $u \leq 10^3$. \\
As the MACHO approaches to the line-of-sight of the \mbox{PSR ,} the impact 
parameter varies (in a first approximation) according to the law :
\begin{equation}
u(t) \, = \, u_{min} \; \sqrt{\,1\,+\,\left( \frac{t}{t_{c}} \right)^2}
\end{equation}
where $u_{min}$ is the minimum impact parameter of the MACHO (figure 1) and 
{\small
\begin{equation}
t_{c} = \frac{R_{E}}{v_{\perp}} \; u_{min} \; \approx 
4.55 \frac{\left[ \frac{u_{min}}{10^2} \right]}{\left[ 
\frac{\beta_{\perp}}{10^{-3}} \right]}
\sqrt{ \left[ \frac{M}{M_{\odot}} \right] \left[ \frac{D_{s}}{4 Kpc} \right] 
\left[ \frac{x_{d}}{0.5} \right]
\left[ 1 - \left( \frac{x_{d}}{0.5} \right) \right]} \;\;\;\; yrs
\end{equation} 
}
is a characteristic time. 
The largest value of $t_{c}$ is 
provided by a mass passing half-way between the observer and the PSR 
($\,x_{d} = 0.5\,$).
A rough estimate of the integral phase residual time for our Shapiro shift 
in a $10\,t_{c}$ characteristic scale (an arbitrary cut-off time chosen 
for experimental data convenience) is :
\begin{equation}
\Delta t \left( u_{min} \right) \, - \, \Delta t \left( 10\,u_{min}\right) 
\approx 45 \left( \frac{M}{M_{\odot}} \right) \;
\; \mu s \;\;\;\; ,
\end{equation}
for large values of $u_{min}$. Detailed residual time evolutions from 
equation 
(1) are described in figures (2) and (3) for deflecting masses $M_{\odot}$
and $10^{-2}\,M_{\odot}$.\\
As we are allowing (on the contrary of the optical gravitational lensing of 
stars) large values of the impact parameter  $u$ we can neglect the first 
vanishing geometrical term in equation (1) and consider only the 
variable term in \mbox{$\Delta t_{grav}$.}\\
The speed of variation of the period $P$ of the PSR is roughly the 
ratio of $\; r_{S} / c \;$ and the characteristic Shapiro phase shift event 
time , namely its time evolution is : 
\begin{equation}
\dot{P} \; = \; \frac{d\,\Delta t_{grav}}{dt} \; = \; 
\frac{d\,\Delta t_{grav}}{d\,u}\;\frac{d\,u}{dt} \, = \, \dot{P}_{0} \, 
F(t)
\end{equation}
where
\begin{equation}
\dot{P}_{0}  = \frac{2\,r_{S}\,\beta_{\perp}}{R_{E}\,u_{min}} \, 
= \frac{1.38 \cdot 10^{-13} \;\left(\frac{\beta_{\perp}}{10^{-3}}\right)
\sqrt{\frac{M}{M_{\odot}}}}{\,\left(\frac{u_{min}}{10^2}\right)
\sqrt{\left(\frac{D_{s}}{4 Kpc}\right)\left(\frac{x_{d}}{0.5}\right)\left[1-
\left(\frac{x_{d}}{0.5}\right)\right]}} \;\;\; s \, s^{-1}
\end{equation}
\begin{equation}
F(t) \, = \, - \, \frac{\left(\frac{t}{t_{c}}\right)}{\sqrt{\left[1 + \left(
\frac{t}{t_{c}} \right)^2 \right]\left[1 + \left(\frac{t}{t_{c}} \right)^2 + 
\left( \frac{2}{u_{min}} \right)^2 \right]}} \;\;\; .
\end{equation}
The considered $\dot{P}_{0}$-value for the present case , when the MACHO is 
half-way , has 
its minimum value. Therefore it defines a characteristic lower bound for 
residual PSRs' period variation.  
The function $F(t)$ has a maximum and a symmetric 
minimum for the values $\, \mp \, t_{max} \,$ , \mbox{where :}
\begin{equation}
t_{max} \, = \, t_{c} \, \left[ 1 + \left( \frac{2}{u_{min}} \right)^2 
\right]^{\frac{1}{4}} \; \approx \; t_{c} \;\;\;\; ,
\end{equation}
(the approximation being valid for large values of $u_{min}$) whose value 
(see equation 4) is nearly $\;4.6 \; \sqrt{\frac{M}{M_{\odot}}} \;\;yr\;$.
In figure (4) the function $F(t)$ is plotted for the values $\;u_{min} = 1\;$ 
and $\;u_{min} \geq 10\;$. It can be seen that as the MACHO approaches to the 
line-of-sight of the PSR (negative values \mbox{of t) ,} the PSR's period 
increases at 
a large rate , but , when $\,u = u_{min}\,$ the growth decreases to zero ; 
the process is time-symmetric.\\ 
Using a double exponential disk model \cite{alcock}, the gravitational lensing 
optical depth can be evaluated. Choosing the galactic centre direction 
\mbox{($l = 0^{\circ} \;,\; b = 0^{\circ}$) ,} the calculation leads to the value 
\begin{equation}
\tau = 2.33 \cdot 10^{-7} \; u_{min}^{2} \, \left( \frac{\rho_{0}}{0.08 
\, M_{\odot} \, pc^{-3}} \right) 
\left( \frac{D_{s}}{4\,Kpc} \right)^{2}
\end{equation}
(being the integrated contribution of matter of the galactic bar component 
very low with respect to the galactic disk one between 4 and 8 Kpc from 
the galactic centre , it has been neglected).
If this value is chosen as a reference one , both larger threshold parameters 
(of at least $u_{min} = 10^{2}$) and the number of available PSRs 
($\,\geq 500\,$)
can make the total optical depth be of the order of unity , i. e. 
a number 
{\small
\begin{equation}
N \, \approx \, \tau \, N_{PSR} \, 
= 1.16 \left[ \frac{u_{min}}{10^2} \right]^{2} 
\left[ \frac{N_{PSR}}{500} \right] \left[ \frac{\rho_{0}}{0.08 \,
M_{\odot} \,pc^{-3}} \right] \left[ \frac{D_{s}}{4\,Kpc} \right]^{2} 
\end{equation}
}
of gravitational host MACHOs can always be found at the same time.\\
A rough estimate of the rate of ``complete'' Shapiro phase shift events is :
\begin{equation}
\Gamma \; \approx \; \tau \, N_{PSR} \; \frac{v_{\perp}}{R_{E}\,u_{min}} \, 
\approx \; 0.51 \;\, \frac{ \left( \frac{u_{min}}{10^2} \right) 
\left( \frac{N_{PSR}}{500}\right)\left(\frac{\beta_{\perp}}{10^{-3}}
\right)}{ \sqrt{\left(\frac{M}{\;M_{\odot}}\right)\left(\frac{D_{s}}{4 Kpc}
\right)}} \;\;\;\; ev \, yr^{-1} \;\; ,
\end{equation}
where in the last member the dependence on $x_{d}$ has been eliminated by an 
exact volume integration. This rate estimate is based on known luminous 
lensing object galactic densities ; in presence of dark matter it may be one 
or two orders of magnitude underestimated.\\ 
A competitive gravitational delay or enhancement in residual PSRs' periods 
may occur when an unbounded object (planet , star , black hole , ...) induce 
a gravitational deflection to the PSR (or the solar system) , leading to 
Doppler shifts (see appendix A1). Another equivalent 
shift occurs also for the presence of binary 
bounded objects near the PSR. Their influence induces a periodic sinousoidal 
residual signal in the PSR's period. Such an effect has been already 
observed at the level of intensity ($\mu s$ over months) comparable to or 
below the one we are considering. However the signatures of such two 
different events are totally 
different (a step function increase or decrease , and a sinousoidal period 
residuals). The corresponding intensity and probability (related to the 
probability to find such an object in the surrounding volumes of the PSR or 
of the solar system) are extremely low with respect to the corresponding
{\bf SPSP} effect , whose probability is defined by much wider and longer 
cylindrical volume.   
\section{CONCLUSIONS AND FIRST EXPERIMENTAL EVIDENCES}
The ({\bf SPSP}) induces a unique characteristic secular variation on the 
PSRs' residual period : its signature leads to a pulse episodic 
bell-shaped time evolution of an order of tens of $\mu s$ for 
one-solar-mass dark objects , well within the present detectability. It will 
never lead to an opposite sign event 
(first a decrease followed by an increase of the period). The pulse  
period variation could not be easily confused by a periodic Doppler shift , 
due to gravitational bounded planets , nor by rarest step increasing (or 
decreasing) Doppler shifts , due to an unbounded planet-star encounter near 
the 
PSR. The characteristic time residual $\Delta t$ (equation 1) event , being 
dominated by a logarithmic function of the adimensional impact parameter 
$u$ , 
has a  wider cross section than gravitational lensing magnification and 
imply 
a larger probability to observe such an event even if at slow time scales. 
The characteristic scattering time is 
\mbox{$\;t_{c} \approx 4.6 \left( 
\frac{u_{min}}{10^2} \right) \left( \frac{\beta_{\perp}}{10^{-3}} 
\right)^{-1} 
\sqrt{ \frac{M}{M_{\odot}} } \;\; yr\;$} 
and the 
corresponding maximum amplitude in the period's derivative evolution 
$\; \left| \dot{P} \right| = \left| F(t) \right| \, \dot{P}_{0} 
\leq 6.9 \cdot 10^{-14} \left( \frac{u_{min}}{10^2} 
\right)^{-1} \left( \frac{\beta_{\perp}}{10^{-3}} \right) 
\sqrt{\frac{M}{M_{\odot}}\,} \;\; s \, s^{-1}\;$ which 
is above or within the present 
instrumental sensitivity and it is even comparable to the present 
characteristic PSRs' period derivatives. The characteristic time evolution 
shown in figure (4) makes the effect observable and well defined. However 
in the present catalogues of PSRs the period derivative $\dot{P}$ has 
already an average positive derivative near $10^{-14}\;s\,s^{-1}\;$. This is 
related to the intrinsic PSR \mbox{spin-down ,} due 
mainly to its luminosity radio 
emission. Therefore the Shapiro positive period derivative in earlier stages 
($\,\dot{P} > 0\,$) might be easily masked by common loss of angular momentum 
of the PSR. On the contrary a negative Shapiro period derivative 
near its maximum amplitude ($\left|F\right| \approx 0.5$) in the second 
phase of the gravitational deflection 
(positive \mbox{times ,} see figure 4) might overcome the positive PSRs' 
period derivative leading to an exceptional 
$\dot{P}$. For $\,u_{min} = 10^2\,$ we must 
expect nearly $\,0.5 \; ev \; yr^{-1}\;$ for $\,M \sim M_{\odot}\,$ at an 
intensity $\;\left| \dot{P} \right| \leq 6.9 \cdot 10^{-14} \;s 
\,s^{-1}\,$ and even more events ($\sim 5 \, ev \, yr^{-1}\;$ for 
$\,M \sim 10^{-2}\,M_{\odot}\,$) but at a 
lower level , $\, \left| \dot{P} \right| \leq 6.9 \cdot 10^{-15} \; s \, 
s^{-1}\,$ , almost on the edge of the characteristic intrinsic positive 
$\dot{P}$. Therefore lower-mass deflectors 
($\;M \leq 10^{-2} \, M_{\odot}\;$) will 
not easily perturbate the median intrinsic positive derivative.
After a first inspection into the up-to-date catalogue of 558 
PSRs \cite{catalogue} we do note the ``anomalous'' negative period 
derivatives shown in 
table 1.
\begin{table}
\caption{PSRs that manifest negative $\dot{P}$.}
\label{negative}
\begin{tabular}{@{}lccl}
PSR & $P$ ($ms$)& $\dot{P}$ ($\;s\,s^{-1}$)& Environment\\
       &     &           &          \\
B 0021-72C & $5.7$ & $-\,4.0 \cdot 10^{-17}$ & C \\
B 1744-24A & $11.5$ & $-\, 1.9 \cdot 10^{-20}$ & B , C\\
B 1813-26 & $59.3$ & $-\,3.0 \cdot 10^{-16}\;$ &  \\
B 2127+11D & $4.8$ & $-\, 1.1 \cdot 10^{-17}$ & C \\
B 2127+11A & $110.6$ & $-\, 2.1 \cdot 10^{-17}$ & C \\
\end{tabular}
\medskip \\
In the last right column the notations mean : C = globular \mbox{cluster ,}
B =  binary PSR 
\end{table}
We note that the absolute $\dot{P}$-values are quite small , but 
assuming an average positive $\dot{P} \sim 10^{-14}\;s\,s^{-1}\;$ we are well 
within our rough predictions ($0.5 \div 5 \;ev\;yr^{-1}$). 
We also note that three 
PSRs (first , second and fourth ones) are found in globular clusters , 
where is more probable to find a star near the line-of-sight. However the 
globular cluster optical depth , even if additional to the external galactic 
one , is not large enough to change significantly the {\bf SPSP} rate 
probability.
Finally we note 
that 85 PSRs has unmeasured $\dot{P}$ , possibly reflecting either intrinsic 
PSR stochastic behaviour or low level 
Shapiro phase shift noise. From figure (4) the maximum variability occurs 
near the maxima or the minima , where $\;\ddot{P} \leq 0\;$. 
Any PSR exhibiting the peculiarity $\;\dot{P} \simeq 0\;$ , 
$\; \ddot{P} \leq 0 \;$ (or in the fast transient linear phase 
when $\dot{P} < 0\;,\; \ddot{P} \simeq 0$ or , finally , in late stages when 
$\;\dot{P} \simeq 0 \; , \; \ddot{P} > 0$) must be carefully kept 
under control.
Their $F(t)$-evolution may lead to the expected double or unique bounce 
from its maximum or from its minimum.\\
A careful analysis and monitoring on $\dot{P}(t)$ evolution and deviation for 
all known PSRs (in particular the 85 whose $\dot{P}$ variability is 
unknown and the above-mentioned five negative ones) may offer a unique 
tool to observe dark matter MACHOs. 
Because of the large impact parameter , the {\bf SPSP} may observe also 
diluited dark molecular clouds which might be spread over a wide radius 
($\;10^{17}\;cm > R > 10^{13}\;cm\;$) and could not be easily observed by 
gravitational lensing magnification. In conclusion we do believe that {\bf 
SPSP} is already acting on PSRs' periods (see also note at proof) and it may 
better
probe the mysterious nature of dark matter in both the galactic disk and 
halo.  
First evidences from a few low $\left( 10^{-16} \div 10^{-17} \right)
\,s\,s^{-1}\,$ negative period derivative seem to favour light MACHO 
candidates of $\;M \leq 10^{-1} \, M_{\odot} \;$ with respect to large ones 
$M \geq M_{\odot}$. Being the only one out of any cluster and any possible 
environmental disturbance , the PSR B1813-16 seems to 
be the best candidate suffering the {\bf SPSP} process.\\
A possible mimic of the {\bf SPSP} may derive from the exchange of angular
momentum from an accreting disk near a PSR , leading to \cite{sha2} 
\begin{equation}
\dot{P} = - 2.33 \cdot 10^{-12} \left( \frac{P}{sec} \right)^{\,2} \left(
\frac{\dot{M}}{10^{17}g\,s^{-1}} \right)^{\frac{6}{7}}  \;\;s\,s^{-1}
\end{equation}
for characteristic PSRs with accreting mass fluxes of 
$\sim 10^{17} \;g\,s^{-1}$. However such a flux would be characteristic of a 
pulsating X-ray source ; moreover the presence of such a disk (even of lower 
flux) near a non-binary object is quite unexpectable : neutron stars are 
assumed to evacuate external volumes after their super-nova birth. The 
alternative (and extremely rare) capture , destruction (by tidal forces)
of a planet would introduce , by keplerian eccentricity , a recognizable 
periodic Doppler modulation. Hypothetical but unknown secular angular 
momentum 
exchange (between external and internal fluids of the PSR) might also lead to 
such a period deviations.\\
The ``artificial'' fine-tuned accretion of a 
circular disk near the PSR \mbox{remains ,} however , the unique alternative 
possibility which we were able to argue. 
\begin{appendix}
\section{GRAVITATIONAL TIME DELAY}
The gravitational time delay term is obtained by integrating along the ray's 
path the potential part of the effective refraction index introduced by the 
MACHO : $\;\Delta t_{grav} = - \, \frac{2}{c^3} \int U dl \, = \frac{r_{S}}{c} 
\int \frac{dl}{r} \;$ \cite{schneider}. The calculation leaded to the 
subsequent more general \mbox{formula :}
\begin{equation}
\Delta t_{grav} = \frac{r_{S}}{c} \, \sum_{i=1}^{4} \, \log \left( 
a_{i} + \sqrt{a_{i}^2 + 1\,} \right) 
\end{equation}  
where
\begin{equation}
a_{1} = \frac{x_{+}}{D_{d}}  \;\;\;\; , \;\;\;\; a_{2} = \frac{x_{+}}{D_{ds}}
\end{equation}
\begin{equation}
a_{3} = \frac{D_{d}}{\left( b + x_{+} \right)} \, \left[ 1 - 
\left( \frac{b}{D_{d}} \right) \left( \frac{x_{+}}{D_{d}} \right) \right] \;\;,
\end{equation}
\begin{equation}
a_{4} = \frac{D_{ds}}{\left( b + x_{+} \right)} \, \left[ 1 -
\left( \frac{b}{D_{ds}} \right) \left( \frac{x_{+}}{D_{ds}} \right) 
\right] \;\; ,
\end{equation}
$x_{+} = \frac{R_{E}}{2} \, \left[ \sqrt{u^2 + 4\;} - u \right]$ being the 
coordinate of the primary image (in the image plane having the origin towards 
the PSR , see figure 1) , 
$b = u R_{E}\;$ the impact parameter and , therefore , $\;\left( b + x_{+} 
\right) = 
\frac{R_{E}}{2} \left[ u + \sqrt{u^2 + 4\,} \right] \;$.\\ 
As long as $u \rightarrow \infty $ , 
$a_{1} \sim \frac{R_{E}}{u\,D_{d}}$ , $a_{2} \sim \frac{R_{E}}{u\,D_{ds}}$ ,
$a_{3} \sim \frac{D_{d}}{u R_{E}} - \frac{R_{E}}{u D_{d}}$ and 
$a_{4} \sim \frac{D_{ds}}{u R_{E}} - \frac{R_{E}}{u D_{ds}}$. 
Therefore , as $u$ tends to infinity $\; \Delta t_{grav} \;$ fades to 
zero like $\; \frac{r_{S}}{c} \, \frac{D_{s}}{b} \;$ (as it is expected).\\
Equation (14) is surely too sophisticated  for our purposes. A simpler 
formula can be deduced from equation (14). In our limit $\;b , R_{E} \ll 
D_{d} \, , D_{ds}\;$ 
\mbox{$\left( u \ll \sim \sqrt{\frac{2\,D_{s}}{r_{S}}}\;\; ,
\;\; r_{S} \ll D_{s} \right)$ ;} therefore  
in this approximation only the third and fourth terms in 
equation (14) are important and , in both $a_{3}$ and $a_{4}$ , the second term 
in the square parenthesis is neglectable with respect to unity. The resulting 
formula is equation (1).    
\subsection{THE DOPPLER SHIFT BY ENCOUNTERS OF NEARBY STARS}
The possible momentum exchange due to an unbounded nearby object (star , 
planet , black hole , ...) in gravitational deflection leads to a tiny 
Doppler shift on the PSR's period. Let us assume a scattering parameter 
$b$ (not to be confused with the above-considered {\bf SPSP} line-of-sight 
impact parameter) by a one-solar-mass object of Schwarzschild radius 
$r_{S_{\odot}}$. The consequent orthogonal momentum transfer leads to an
adimensional velocity exchange on the 
primordial PSR's motion is \mbox{\cite{df,df2} :} 
\begin{equation}
\langle \Delta \beta_{PSR} \; \rangle \; \; \simeq \; \;
\left\langle \, \frac{r_{S_{\odot}}}{b} \; 
\frac{\left(1+\beta^{\,2}_{\odot}\right)}{\beta_{\odot}} \, 
\right\rangle \; 
\leq \;\,3.0 \, \cdot 10^{-9} \left( \frac{b}{0.1\,l\,y} \right) 
\left( \frac{\beta_{\odot}}{10^{-3}} \right)^{-1} 
\end{equation}
where $\beta_{\odot}$ is the adimensional relative velocity between the PSR 
and the object. The corresponding average period derivative 
$\left\langle \dot{P} \right\rangle_{Dp}$ during the whole time $\Delta T$ 
of the gravitational deflection is
\begin{equation}
\left\langle \dot{P} \right\rangle_{Dp} \;\; \cong \;\; 
\frac{\,\Delta P \, }{\Delta T} \;\; \simeq \;\;
\frac{\;\left( \frac{r_{S_{\odot}}}{b\,\beta_{\odot}} \right)\;}{\;
\left( \frac{b}{\beta_{\odot}\,c} \right)\;} \;
\approx \; 10^{-18} \cdot \left( \frac{P}{sec} \right) \; 
\left( \frac{b}{0.1\,l\,y} \right)^{-2} 
\left( \frac{M}{\;M_{\odot}} \right) \;\; s \; s^{-1} \;.
\end{equation}
This value which already assume a high and unprobable star density near
a PSR ($\, \geq 10^{\,5} \; pc^{-3}\,$) is negligible for 
most PSR's periods with respect to the characteristic {\bf SPSP} delay.\\
In addition to the orthogonal momentum transfer , the parallel momentum 
transfer , in analogy to the electromagnetic bremsstrahllung , can reproduce 
also a bell-like {\bf SPSP} time delay. 
If we observe the PSR with the line of sight in the same direction of 
propagation of the nearby crossing object (MACHO , Juppiters ...) , because of 
the exchange of parallel momentum ,  
the PSR is sent away and back by , \mbox{roughly ,} a distance
$\; \Delta l_{max} \approx \frac{G\,M}{\beta_{\odot}^2 \,c^2} \;$ in a 
characteristic time $\; \frac{b}{\beta_{\odot} \,c} \;$ , producing a 
bell-like time delay whose maximum value is 
\begin{equation}
\Delta t_{\| max} \; \approx \; \frac{r_{s}}{c} 
\frac{1}{\beta_{\odot}^{2}} \;\;\;\;. 
\end{equation}   
Therefore the final figure will be a generic combination of the two effects 
(bell-like for parallel and step-like for orthogonal momentum transfer). 
We note that such bell-like PSR's period evolution may be sign-reversed 
because of the time-reversibility of the PSR-object scattering.
In conclusion , out of a very rare geometrical configuration (in 
line of sight of the scattering direction) , the process should be in general 
distinguishable from our unique positive bell-like {\bf SPSP} evolution.
\section*{NOTE ADDED IN PROOF}
We acknowledge the referee for letting us notice a previous article 
\cite{larch} where the {\bf SPSP} have been applied to the residual period 
evolution of the PSR B0525+21 ($D_{s} = 2.3 \, Kpc$ , 
\mbox{$l \approx 183^{\circ}$ ,}
$b = 6.9^{\circ}$). The authors found out a bell-shaped evolution which is 
similar to 
the {\bf SPSP} effect. However , even if their formulation of the integral 
phase residual time agrees with our 
\mbox{$\;\Delta t (u) - \Delta t (u_{min}) 
\approx - \frac{r_{S}}{2c} \log \left[ 1 + \left( \frac{t}{t_{c}} \right)^2 
\right]\;$} 
\mbox{(for large $u$) ,} they did not discuss the optical depth for 
such a gravitational lensing event. From the fitting parameters of the 
deflecting mass ($\;M \approx 330\,M_{\odot}\;$) and characteristic evolution 
time ($t_{c} \approx 0.32\,yr$) , equation (4) allows a minimum impact 
parameter of 
{\small
\begin{equation}
u_{min} = \frac{0.51}{\sqrt{\left(\frac{x_{d}}{0.5}\right)
\left[1 - \left(\frac{x_{d}}{0.5}\right)\right]}} \; 
\left(\frac{\beta_{\perp}}{10^{-3}} \right) \;\;\;\;.
\end{equation}}
As the most of the relative contribute 
$\;\frac{1}{\tau}\,\frac{d\,\tau}{d x_{d}}\;$ 
to the optical depth is provided when the deflector is nearly half-way the 
distance $D_{s}$ , we conclude that $\;u_{min} \leq 1\;$. The theoretical 
optical depth for the PSR B0520+21 (nearly in the galactic anticentre 
direction) is $\;\tau \leq 2.11 \cdot 10^{-8} \;$. Due to this 
extremely low probability as well as the absence of the unavoidable twin 
lensing image ($u_{min} \leq 1$) , we were forced to reconsider a slightly 
more probable 
explanation of the Doppler shift encounter (see appendix A1). Contrary to our 
earliest 
considerations for the orthogonal momentum transfer (see equations 18 , 19) ,
the parallel momentum transfer (equation 20) can reproduce a bell-like 
{\bf SPSP} delay (with some additional step-like pattern).\\
We have imagined a free-free gravitational scattering and we noted that , if 
the largest delay must be $\Delta t_{\| max} \approx 32 \, ms\;$ and if it is 
reached in nearly $
\frac{b}{\beta_{\odot}\,c} \approx 5 yr$ , it might be produced by a mass of 
$M 
\approx 10^{-3} \left( \frac{\beta_{\odot}}{10^{-3}} \right)^2 \, M_{\odot}$ 
with an impact parameter $b \approx 4.5 \cdot 
10^{15}\, \left( \frac{\beta_{\odot}}{10^{-3}} \right)\;cm$. However a rough 
estimate of the probability for such a nearby encounter should be 
$\approx 5 \cdot 10^{-7} \div 10^{-6} \;$ , respectively for realistic luminous 
or dark matter assumption ; still a small and puzzling value.
\end{appendix}
\newpage
\begin{figure}
\begin{picture}(410,200)(0,0)
\put(145,67){\circle*{6}}
\put(155,67){\makebox(0,0)[l]{MACHO}}
\put(405,102){\circle{10}}
\put(405,92){\makebox(0,0)[t]{OBS.}}
\put(400,102){\line(1,0){10}}
\put(405,97){\line(0,1){10}}
\multiput(15,102)(10,0){40}{\line(1,0){6}}
\put(10,102){\circle{10}}
\multiput(145,170)(-4,1){34}{\circle*{1}}
\put(10,112){\makebox(0,0)[b]{PSR}}
%
\put(145,170){\line(4,-1){256}}
\put(145,170){\line(-2,-1){130}}
\put(137,67){\line(0,1){35}}
\put(134,67){\line(1,0){6}}
\put(134,102){\line(1,0){6}}
\multiput(145,72)(0,10){9}{\line(0,1){5}}
\put(145,102){\vector(0,1){105}}
\put(147,172){\makebox(0,0)[lb]{\large $x_{+} = 
     \frac{R_{E}}{2} \left[ \sqrt{u^2 + 4} - u \, \right]$}}
\put(143,105){\makebox(0,0)[br]{\small $0$}}
\put(132,82){\makebox(0,0)[r]{$b = u \, R_{E}$}}
\put(140,210){\makebox(0,0)[bl]{\large $x$ , image plane}}
\put(145,45){\line(1,0){260}}
\put(145,42){\line(0,1){6}}
\put(405,42){\line(0,1){6}}
\put(270,42){\makebox(0,0)[t]{\large $D_{d} = x_{d}\,D_{s}$}}
\put(145,45){\line(-1,0){135}}
\put(10,42){\line(0,1){6}}
\put(74,42){\makebox(0,0)[t]{\large $D_{ds} = \left( 1 - x_{d} \right) D_{s}$}}
\put(10,15){\line(1,0){395}}
\put(10,12){\line(0,1){6}}
\put(405,12){\line(0,1){6}}
\put(180,12){\makebox(0,0)[t]{\large $D_{s}$}}
\end{picture}
\caption{Relative position of source , MACHO , observer and primary image. As
$u \rightarrow \infty$ , $x_{+} \rightarrow 0$ and the image corresponds to
the source. $v_{\perp}$ accounts for the relative PSR-MACHO-observer orthogonal 
motion velocity.}
\end{figure}
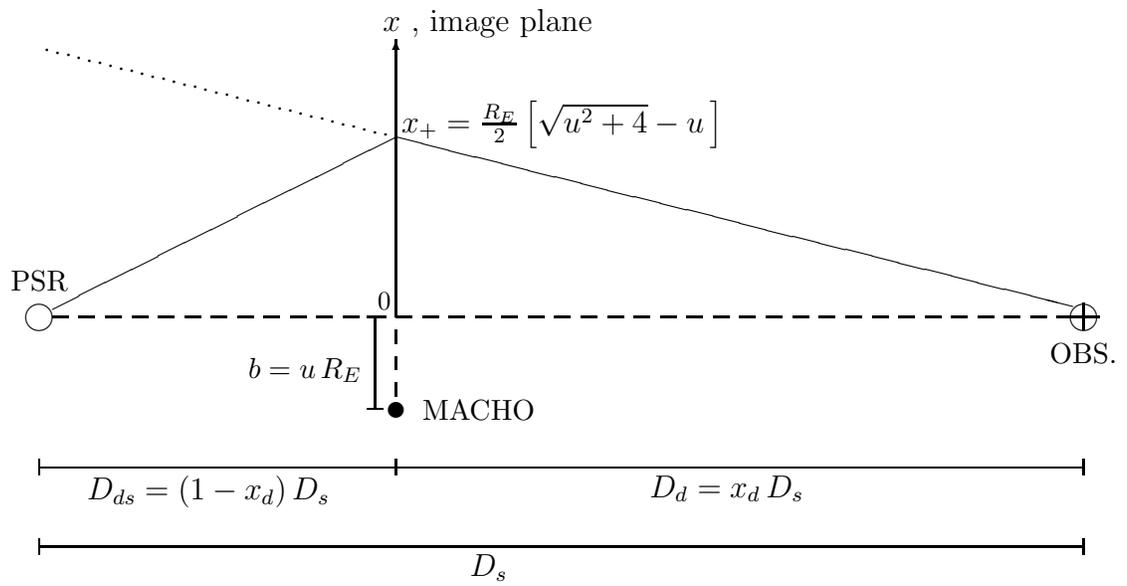
\end{document}